# Mesoscopic Phase Coherence in a Quantum Spin Fluid.

Guangyong Xu[1,2], C. Broholm[1,3], Yeong-Ah Soh[4], G. Aeppli[5], J. F. DiTusa[6], Ying Chen[1,3], M. Kenzelmann[1,3], C. D. Frost[7], T. Ito[8], K. Oka[8], H. Takagi[8,9].

[1]*Department of Physics and Astronomy, The Johns Hopkins University, Baltimore, Maryland 21218, USA*

[2]*Condensed Matter Physics and Materials Sciences Department, Brookhaven National Laboratory, Upton, New York 11973, USA*

[3]*NIST Center for Neutron Research, National Institute of Standards and Technology, NIST, Gaithersburg, Maryland 20899, USA*

[4]*Department of Physics and Astronomy, Dartmouth College, Hanover, NH 03755, USA*

[5]*London Centre for Nanotechnology and Department of Physics and Astronomy, University College London, 17-19 Gordon Street, London, WC1H OAH  UK*

[6]*Department of Physics and Astronomy, Louisiana State University, Baton Rouge, Louisiana, 70803, USA*

[7]*ISIS Facility, Rutherford Appleton Laboratory, Chilton, Didcot OX11 0QX, UK.*

[8]*National Institute of Advanced Industrial Science and Technology, Tsukuba, Ibaraki 305-8562, Japan*

[9]*Department of Advanced Materials Science, Graduate School of Frontier Sciences, University of Tokyo, Kashiwa, Chiba 277-8561 , Japan*

**Mesoscopic quantum phase coherence is important because it improves the prospects for handling quantum degrees of freedom in technology. Here we show that the development of such coherence can be monitored using magnetic neutron scattering from a one-dimensional spin chain $Y_2BaNiO_5$, a quantum spin fluid where no classical, static magnetic order is present. In the cleanest samples, the quantum coherence length is 20 nm, almost an order of magnitude larger than the classical antiferromagnetic correlation length of 3 nm. We also demonstrate that**

**the coherence length can be modified by static and thermally activated defects in a quantitatively predictable manner.**

A wave is deemed coherent when it is perfectly periodic in space and time. Quantum mechanical phase coherence means that the corresponding wavefunction is actually a coherent wave. Typically, though, classical disorder limits quantum phase coherence. The occasions where this is not the case and the quantum phase coherence length far exceeds that given by classical considerations remain a major theme of physics. The longstanding examples are superconductivity and superfluidity where macroscopic quantum phase coherence (*1*) has been revealed by persistent current and Josephson junction experiments (*2,3*). The effects of phase coherence have also been seen recently in the incompressible quantum fluid phase of the two-dimensional electron gas (*4*) and in a laser-cooled Bose-Einstein condensate (*5*). Curiously, while magnetism itself is a quantum mechanical phenomenon, originating from the shell model for the atom, mesoscopic (in the sense of extending over distances in excess of 10 nm) quantum coherence in the absence of classical, static order has not been explicitly demonstrated for a magnet. We show that the development of such coherence can be readily monitored using magnetic neutron scattering from a one-dimensional spin chain, in the form of an oxide of nickel, and that it is limited by static and thermally activated defects.

One of the best known magnets which remains disordered due to quantum fluctuations is a chain of integer spins coupled antiferromagnetically (i.e. favoring antiparallel alignment) to their neighbors (*6,7*). The two-spin correlation function $\langle \mathbf{S}_j \cdot \mathbf{S}_{j'} \rangle$ decays exponentially with large distances $|j - j'|$ between spins $\mathbf{S}_j$ and $\mathbf{S}_{j'}$ in the chain as in a one-dimensional liquid; whereas in a classical magnet static correlations persist over macroscopic length scales when the system orders. Even so, theoretical work has shown that the ground state is coherent in the sense that the more complex, 'string' correlation function (*8*), $O(j-j')=\left\langle S_j^z \exp\left[i\pi\sum_{j<k<j'} S_k^z\right] S_{j'}^z \right\rangle$, tends to a constant (*9*) for $j-j' \to \infty$. A simple way to understand the string order is to look at the spin chain in terms of the spin projection quantum numbers $S_z = -1, 0$ and $+1$ at individual sites. Correspondingly, for S=1, the basis states are $|\pm 1\rangle$, which have clear classical analogs and can be used to build a perfect Néel state $|....-1,+1,-1,+1,-1,+1.....\rangle$ for the chain, and $|0\rangle$, which does not. String order is carried by states where sites with $S^z=0$ are inserted into the perfect Néel state (see Fig. 1B). A good approximation to the ground state wavefunction for the Heisenberg S=1 chain is a coherent superposition of all such states with a weighting factor that decreases exponentially with the number of $|0\rangle$ sites (*10*). In general, long range quantum coherence is believed to exist in all one-dimensional systems with spectral gaps (*11*).

For the spin-1 chain, the lowest excited state above the ground state entails inserting a defect into the string order. The defect can propagate throughout the lattice (Fig. 1C) until it encounters another defect, and so will have a mean free path limited by either static defects, such as chain terminations (*12,13*) (Figs. 1E and 1F) or other moving, thermally excited defects of the same type (Fig. 1D). Thus, the coherence of the ground state should be manifested in the mean free path of its triplet excitations. By imaging these excitations and their mean free path using inelastic neutron scattering, we can establish the perfection, or quantum coherence, of the ground state.

Neutron scattering measures the Fourier transform $S(q,\omega)$ in space and time of the magnetic correlation function $\langle \mathbf{S}_j(t) \cdot \mathbf{S}_{j'}(t') \rangle$. The experiments were conducted on single crystals of $Y_2BaNiO_5$, which contain an ensemble of spin-1 chains formed by the O-Ni-O backbone of corner-linked $NiO_6$ octahedra (*14,15*). The system behaves as a quantum spin liquid due to the one-dimensional nature of the magnetic interactions. That this liquid is not a conventional classical liquid, however, is clear from Fig. 1A, which shows the full dynamical structure factor $S(q,\omega)$, multiplied by energy transfer $\hbar\omega$, for $Y_2BaNiO_5$. Instead of broad over-damped excitations, which soften toward zero energy as the ordering vector ($q = \pi$ in this case) for the corresponding classical solid is approached; a gapped band of well-defined modes dominates the data. The lowest energy mode appears at $q = \pi$, and its energy is termed the Haldane gap *(6,7)*. While $S(q,\omega)$ from chemically doped $Y_2BaNiO_5$ is qualitatively similar there are important differences (Fig. S1). Our paper focuses on characterizing the mean free path and energetics of the lowest energy Haldane gap mode.

The temperature(*T*)-dependent equal-time spin correlation length $\xi_0$, is shown as open circles in Fig. 2. $\xi_0$ is inversely proportional to the peak width of the equal time spin correlation function, $S(q) = \sum_{jj'} \langle \mathbf{S}_j \cdot \mathbf{S}_{j'} \rangle \exp(iq(j-j'))$, which in the $T \to 0$ limit displays directly the magnetic correlations built into the ground state. S(*q*) is the magnetic analog of the structure factor measured by X-rays for ordinary liquids. Represented by the open circles in the insets A and B of Fig. 2, $S(q)$ has a maximum at $q = \pi$, indicating that the underlying classical solid is antiferromagnetic (AFM) where each spin points in a direction opposite to its two nearest neighbors. The peak width however does not vanish even as $T \to 0$, and the correlations between spins remain liquid-like for $T \to 0$, as anticipated by Haldane *(6,7)* and seen experimentally in several S=1 chain compounds *(14,16-18)*. The dash-dotted line in Fig. 2 is the prediction of the quantum nonlinear σ-model *(19)* which is only valid for $T < \Delta/k_B$, where Δ is the Haldane gap.

The solid circles in the inset of Fig.2 are the magnetic structure factor measured as a function of *q* with the energy transfer fixed, at the Haldane gap energy for inset A and B; and at 36 meV for inset C. The peak approaches a resolution-limited delta-function at low temperature (inset A). Because we have tuned to the gap energy, the process probed is the creation of a triplet wave packet at rest. The intrinsic width of this peak is therefore inversely proportional to the distance over which the triplet retains quantum phase coherence. The data shows that $\delta q \approx 0.02\pi$ implying phase coherence over approximately 50 lattice units $a$ – much longer than the measured correlation length for antiferromagnetic order, $\xi_0$. The experiment thus provides direct evidence for mesoscopic coherence in the spin-1 chain. The peak broadens significantly on warming (inset B) and doping by chemical impurities (inset C), indicating that quantum coherence in this system requires low temperatures and perfect crystallinity. The two length scales obtained from the two types of $q$-scans vary differently with temperature. At high $T$, the lengths converge to nearly the same value. In contrast, as $T \to 0$, the mean free path for the Haldane gap excitation $\xi$ nearly diverges even while the AFM length $\xi_0$ approaches a finite value. The AFM length is the distance between defects in the AFM order, which at $T$=0, is simply the typical distance between sites with $S_z = 0$. Note that such defects do not disrupt string order. On warming, excitations in the form of ferromagnetic neighbor pairs with $S_z$ both equal to +1 or -1 will also appear, and these not only disrupt antiferromagnetism but also the underlying string order. Classical (gapless) AFM chains (*20-22*), such as the Mn$^{2+}$ (S=5/2) chain compound TMMC (*20*), behave very differently from Y$_2$BaNiO$_5$ in the sense that a single distance, namely the equal –time correlation length $\xi_0$, sets the scale for both the static and dynamic correlations.

Detailed analysis of thermal conductivity provides estimates of the magnetic mean free path in various spin chains and ladders (*23*). The results (*24*) for Y$_2$BaNiO$_5$ differ considerably from our direct measurements in the sense that they pass through a strong peak of *15a* at 100K and rapidly decrease below that, with a value of *7a* at 70K, the lowest temperature at which there was confidence in the measurements. The discrepancy with our results is not surprising given the indirect nature of the thermal conductivity measurements as well as the fact that the lower temperature magneto-thermal properties of S=1 chains are dominated by the effective spins of chain ends rather than the triplet excitations of the entire chain segments.

To quantify the effects of chemical impurities on quantum phase coherence, consider an excited triplet wave packet on the chain. At low temperatures, the population of such wave packets is relatively low and they can travel almost freely before encountering a physical impurity, e.g., a chain end caused by a chemical defect. Correspondingly, the mean free path is on average half of the spin chain length. To obtain an independent measure of the latter, we recall specific heat work (*25*) on $Y_2BaNiO_5$ which shows the residual S=1 degrees of freedom expected for chains containing odd numbers of Ni atoms (*10,13,26*). The outcome is a Curie-like dc magnetic susceptibility $C/T$, which we have measured for the nominally pure material using SQUID magnetometry. The resulting Curie constant, *C*, gives an average impurity concentration of ~1.5%, or a mean distance λ of ~70 lattice spacings between chain breaks, which is, within error, consistent with twice the zero-*T* coherence length measured in our inelastic neutron scattering experiments. The same is true for our deliberately doped (with Ca and Mg) samples. The coherence length extracted from samples with nominal Ca doping of 4% and 10% ($Y_{2-x}Ca_xBaNiO_5$), and Mg doping of 4% ($Y_2BaNi_{1-x}Mg_xO_5$), are 9(1), 5(1), and 10(1) Ni-Ni spacings *a*, respectively, consistent with the scenario suggested above, for which $2\xi/a = 1/x$.

With increasing temperature and therefore triplet population, the mean free path will be limited by both thermally created defects and chain ends. In keeping with the physical picture developed above, the coherence length should approximately match half the inverse density of all (chemical and thermal) defects $\xi/a = (x + \rho(T))^{-1}/2$. Here $\rho(T) = 3\sqrt{\frac{k_B T \Delta}{2\pi v^2}} \exp(-\Delta/k_B T)$ is the thermal triplet density (*27*) and *v*=70 meV the triplet velocity (*14*). The corresponding solid line in Fig. 2 shows that this model provides an excellent description of the *T*-dependent quantum coherence length.

Another consequence of the above picture should be a finite temperature blue shift as triplets confine each other into boxes. In other words, quantum mechanics acts not only to produce the singlet Haldane ground state and the associated gap to triplet excited states, but also gives quantum particle-like properties to the excited states, including a kinetic quantum confinement contribution. Instead of being created with energy $\Delta$, the triplet is created with an additional kinetic energy given by the curvature of the dispersing mode shown in Fig. 1A, associated with a quasi-particle of non-zero momentum $q \sim 1/\xi$. Such a quantum contribution is found - remarkably - on warming as well as doping, where the quantum confinement is due to fixed boundaries. Experimentally, others have also noticed the thermally-induced blue shift for $Y_2BaNiO_5$ (*28*) as well other Haldane systems (*17,29,30*).

Fig. 3 A and B show the temperature dependence of the mean triplet energy ($\Delta$), and the half-width at half-maximum HWHM ($\hbar\Gamma$) of the triplet, assuming Gaussian or Lorentzian spectral functions, respectively. The measured Haldane gap (see also Fig.S2) moves to higher energies on warming, which is exactly the thermal blue shift anticipated. The energy width of the triplet excitation in the $T \rightarrow 0$ limit is resolution-limited, imposing an upper bound of $\hbar\Gamma=0.1$ meV on the excited state relaxation rate. This means that the lifetime of the triplet is $>10^{-10}$ seconds, implying a quality factor for the triplet oscillations $Q=\Delta/\hbar\Gamma>100$. This represents the largest (by an order of magnitude) bound on $Q$ reported to date for an S=1 chain. When we introduce excited states through warming, the HWHM increases, thus providing evidence for an increasing triplet relaxation rate, which goes hand in hand with the reductions in mean free path discussed above.

While the quantum nonlinear σ-model (*19*) provides a reasonable prediction for the spin correlation lengths, it severely over-estimates the thermal blue shift (the dash-dotted line in Fig. 3 A). A more successful approach exploits Monte Carlo calculations (*31*) which provide the *T*=0 gap energy due to finite chain confinement. By using twice the coherence length $\xi$ obtained in Fig. 2 as the size of the confinement "box" (see top scale in Fig. 3A), the gap energies can be calculated and plotted as the blue solid line in Fig. 3. The agreement between experimental data and the parameter-free calculation is truly remarkable. It is also worthwhile to note that introducing chemical impurities is another approach to creating the same quantum confinement effect. Average gap energies for 4% and 10% Ca doped and 4% Mg doped $Y_2BaNiO_5$ are extracted from measurements done at base temperature and plotted as diamonds in Fig. 3 A, which agree very well with the calculated gap energies based on $2\xi/a = 1/x$.

While we have described all of our *T*-dependent data in terms of a very simple model whose underlying driver is thermal activation across a gap to triplet formation, it is important to realize that this model is largely phenomenological. Fortunately, there is a rigorous formulation, with similar conclusions, due to Damle and Sachdev (*27*), whose finite temperature theory is based on a semi-classical treatment of scattering between triplet excited states with a temperature-independent quadratic dispersion relation. While this parameter-free theory theory does not account for the thermal blue shift, it determines the triplet lifetime and leads to the phenomenological expression for ξ in the clean limit (*x*=0), in impressive agreement (see the dashed lines in Figs. 2 and 3 B) with the data.

Our experiments have shown that while quantum spin chains with a gap have a finite equal time spin correlation length, the coherence of excited states at low temperature is limited only by quenched disorder. At finite temperatures the coherence length decreases and the excited states acquire a finite mean free path due to collisions with other excitations. Both effects are quantitatively accounted for by a semi-classical theory of triplet wave packets scattering from each other and from defects in the sample. This demonstration of mesoscopic phase coherence in a magnet adds a fifth example – the other four are superconductors, superfluids, fractional quantum Hall states, and optically confined Bose Einstein Condensates – to the list of systems where quantum phase coherence has been demonstrated in the absence of classical order.

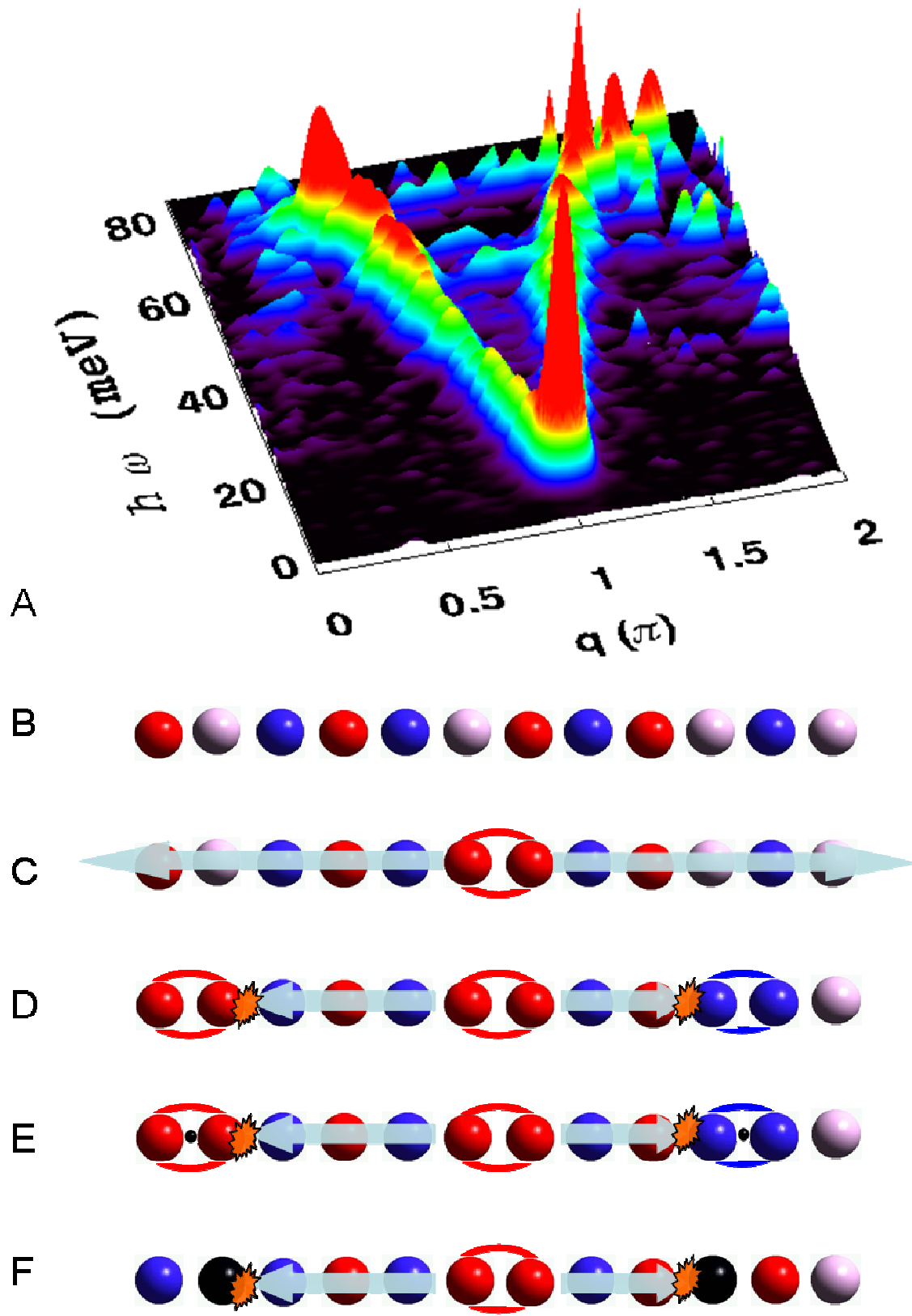

11Figure 1. A. Map of the dynamic spin correlation function multiplied by energy for the one dimensional spin-1 chain $Y_2BaNiO_5$ at low temperature ($T$=10 K). B-F: Diagrams of the string order in the S-1 chain. The red, blue, and grey spheres represent sites with $S^z$ of 1,-1, 0 respectively. Chemical impurities are represented by black spheres (Mg). B: S=1 chain at $T$=0 and no chemical impurity; C: one triplet wave-packet propagating on the chain; D: at $T$>0, triplet wave-packets act as thermal defects and confine each other into boxes; E: box confinement by Ca impurities, which introduce holes (indicated by small black spheres) into the chain, and cause neighboring spins to form ferromagnetic pairs; F: box confinement by Mg impurities, which remove Ni spins from the chain.

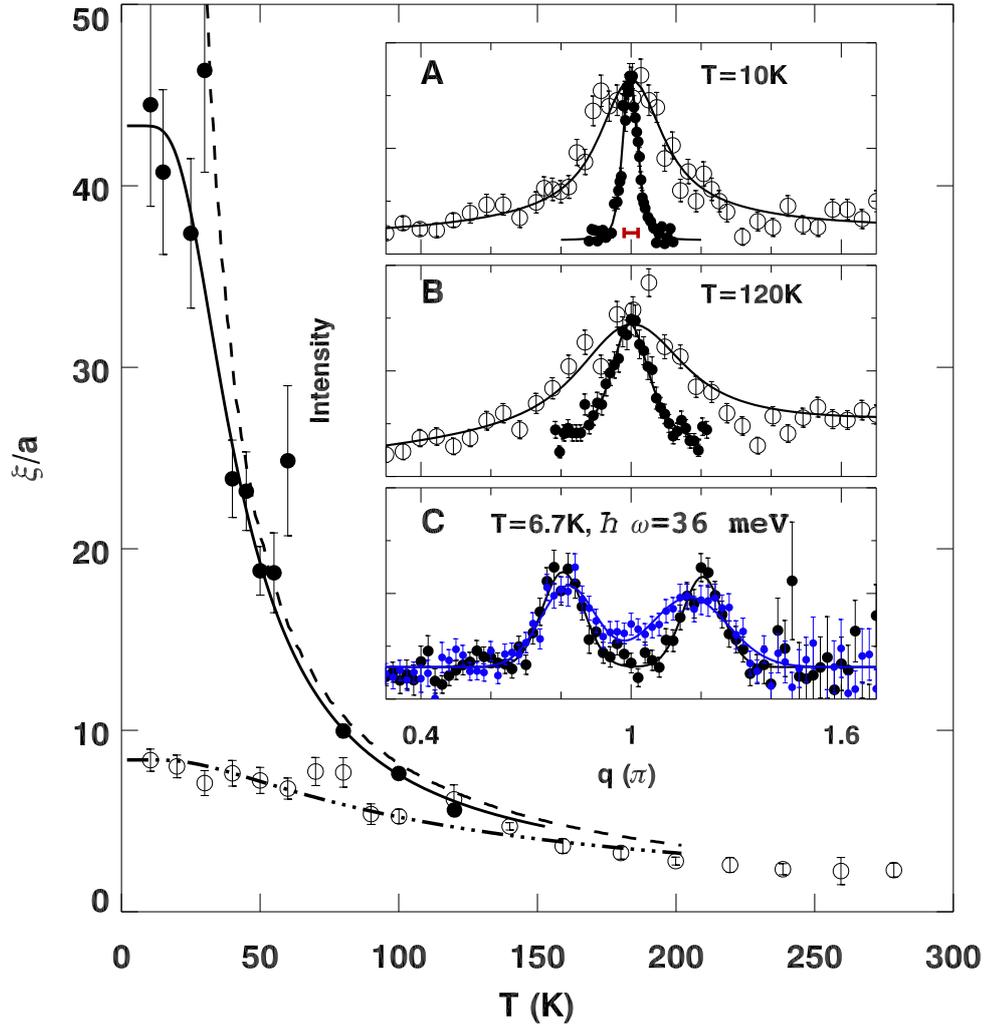

Figure 2. Equal time correlation length $\xi_0$ (open symbols) and magnetic phase coherence length $\xi$ (closed symbols) for the one dimensional spin-1 chain $Y_2BaNiO_5$, as measured with inelastic magnetic neutron scattering (inset) Lines are theoretical results described in the text. Inset (A) shows low temperature structure factor data obtained for the equal-time spin correlations ($\hbar\omega$>8 meV) (open circles), and the bottom of the Haldane continuum at $\hbar\omega$=7.5 meV (closed circles, where instrument resolution is shown as the horizontal bar) in a nominally pure sample. Fits to such data produced the main frame data points as described in the supplement. Inset (B) shows higher temperature data where the coherence length approaches the equal-time correlation length. Inset (C) shows data acquired at a fixed energy transfer (36 meV) for the nominally pure (black symbols) and Ca doped ($Y_{2-x}Ca_xBaNiO_5$ x=0.1) samples (blue symbols).

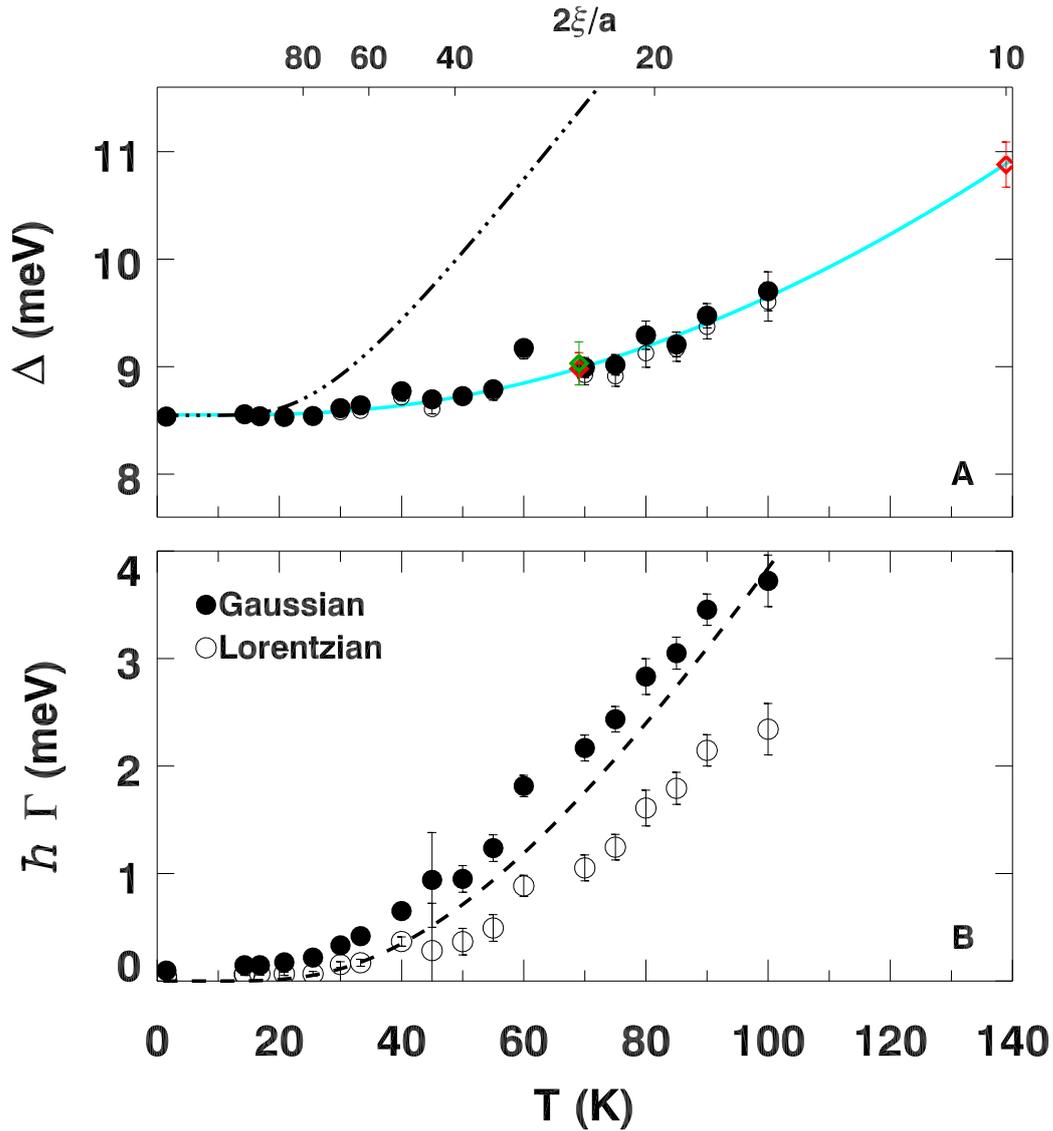

Figure 3. A: Haldane gap energy, and B: half width of the excited triplet mode measured at different temperatures for Y$_2$BaNiO$_5$. The open circles and closed circles are results from assuming Lorentzian and Gaussian spectral functions, respectively, which the data do not distinguish. The lines in A are model calculations described in the text. The dashed line in B is a parameter free theoretical result (*27*). The diamonds are gap energies derived from measurements on Ca and Mg doped samples, and the height of blue bars reflects the uncertainty of the measured coherence lengths in the doped

compounds. The diamonds are gap energies derived from measurements at base temperature on Ca (red) and Mg (green) doped sample (concentration $x$) and placed according to $2\xi/a = 1/x$ of the top scale in (A), which indicates the effective thermal chain lengths for a pure sample (twice the coherence length from dashed line in Fig. 2) resulting from thermal defects.

Technologies program of RCUK. Work at JHU, NIST, and LSU was supported by the U.S. National Science Foundation, while that at BNL was supported by Office of Science, U.S. Department of Energy.